\documentclass[aps,epsf,
twocolumn,prl]{revtex4-1}
\newcommand{\expec}[1]{\left < #1\right >}

\usepackage{amsmath}
\usepackage{amsfonts}
\usepackage{graphicx}
\usepackage{amsthm}
\usepackage{subfigure}
\usepackage{hyperref}
\usepackage[usenames]{color}

\usepackage{rotating}

\usepackage{braket}

\renewcommand{\v}[1]{\mathbf{#1}}

\newcommand{\tworow}[2]{\genfrac{}{}{0pt}{}{ #1}{#2}}

\newcommand{\lp}{\left ( }
\newcommand{\rp}{\right ) }
\newcommand{\lb}{\left [ }
\newcommand{\rb}{\right ] }

\newcommand{\beq}{\begin{eqnarray*}}
\newcommand{\eeq}{\end{eqnarray*}}

\newcommand{\be}{\begin{eqnarray}}
\newcommand{\ee}{\end{eqnarray}}

\newcommand{\mc}{\mathcal}

\def\lsim{\mathrel{\rlap{\lower4pt\hbox{\hskip1pt$\sim$}}
    \raise1pt\hbox{$<$}}}                

\def\gsim{\mathrel{\rlap{\lower4pt\hbox{\hskip1pt$\sim$}}
    \raise1pt\hbox{$>$}}}                

\usepackage{subfiles}

\begin{document}

\title{High temperature thermodynamics of fermionic alkaline earth atoms in optical lattices}

\author{Kaden R.~A. Hazzard$^{1,2}$} \email{kh279@cornell.edu}
\author{Victor Gurarie$^2$}
\author{Michael Hermele$^2$}
\author{Ana Maria Rey$^{1,2,3}$}
\affiliation{$^{1}$ JILA, University of Colorado, Boulder, Colorado 80309-0440, USA}
\affiliation{$^{2}$ Department of Physics, University of Colorado, Boulder, Colorado 80309-0440, USA}
\affiliation{$^{3}$ NIST, Boulder, Colorado 80309-0440, USA}

\begin{abstract}
We calculate experimentally relevant properties of trapped fermionic alkaline earth atoms in an optical lattice, modeled by the $SU(N)$ Hubbard model. Our calculation is accurate when the temperature is much larger than the tunneling rate, similar to current regimes in ultracold atom experiments.  In addition to exploring the Mott insulator-metal crossover, we  calculate final temperatures achieved by the standard experimental protocol of adiabatically ramping from a non-interacting gas, as a function of initial gas temperature and final state lattice parameters.  Of particular experimental interest, we find that increasing $N$ gives substantially \textit{colder} Mott insulators, up to more than a factor of five for relevant parameters.  This cooling happens for all $N$,  fixing the initial entropy, or for all $N \lsim 20$ (the exact value depends on dimensionality), fixing the initial temperature.
\end{abstract}


\maketitle

\textit{Introduction.---}The recent achievement of Fermi degeneracy and Bose-Einstein condensation in ultracold alkaline earth atoms~\cite{takasu_spin-singlet_2003}
 opens up great opportunities   in  quantum information processing~\cite{daley_quantum_2008},
  quantum simulations~\cite{gorshkov_two-orbital_2010}, atomic clocks experiments~\cite{ido_recoil-free_2003}
 and other precision measurements~\cite{kotochigova_prospects_2009}. One fundamental property of fermionic alkaline earth atoms  is their intrinsic  SU($N=2I+1$)-symmetry in the  nuclear spin ($I$) degrees of freedom, while bosonic alkaline earth isotopes   in contrast have even-even nuclei and necessarily  $I=0$~\cite{hyde:nuclear}. Fermionic alkaline earth atoms  loaded in an optical lattice  are described by   the SU($N$) Hubbard model where $N$ can be varied  for a single isotope from 2 to $2I+1\leq  10$  by  selectively populating hyperfine levels.   Cold atoms realizations of this model open up a range of exciting and exotic physics relevant to condensed matter: it is a simple limit of multiorbital models describing transition metal oxides, is important in theoretical generalizations of the Fermi-Hubbard model, and displays (frequently exotic) phenomena such as possible antiferromagnetism, superconductivity,nematic order, valence bond, and spin liquid phases~\cite{gorshkov_two-orbital_2010,cazalilla_ultracold_2009,hermele_mott_2009}.
As a first step towards reaching the low temperatures necessary to observe these states, here we study the SU(N) Hubbard model's high temperature Mott-metal crossover.
One particularly interesting finding is that increasing $ N$ can lead to more than a  five-fold decrease in temperature compared to $N = 2$.

We calculate density and entropy profiles of lattice alkaline earth atoms to second order in $t/T$, the tunneling rate over the temperature (see Eq.~\eqref{eq:SUN-fermi-hubb}).  This calculation is accurate for $T\gg t$, regardless of the on-site interaction $U$. This includes the ``unquenched Mott insulator (MI)" regime $t\ll T \ll U$ that has been realized in $SU(2)$ alkali gases~\cite{joerdens_mott_2008}.
For the $SU(2)$ spin-$1/2$ case, sophisticated,  numerically intensive algorithms have yielded series to tenth order~\cite{oitmaa_series_2010}. For properties in the experimental regime,  $T\gsim t$ in three dimensions, the second order expansion  agrees quantitatively ($\lsim 1\%$ error) with high order expansions and dynamical mean field theory~\cite{jaerdens_quantitative_2010}. We also calculate final temperatures achieved by standard experimental adiabatic ramping protocols.  Excitingly, our calculations show that applying the same protocols as in  $SU(2)$ experiments will generate colder,
less compressible MI states  as $N$ increases, up to $N\sim 10$,
with the most favorable $N$ depending on dimensionality.

Alkaline earth atoms in deep optical lattices are well-described by the $SU(N)$ Fermi-Hubbard model~\cite{gorshkov_two-orbital_2010}
\be
H &=& -t \sum_{\langle ij \rangle, \alpha} f^\dagger_{\alpha,i} f_{\alpha,j} + \frac{U}{2} \sum_{i,\alpha,\alpha'} f^\dagger_{\alpha,i}f^\dagger_{\alpha',i}f_{\alpha',i}f_{\alpha,i} \label{eq:SUN-fermi-hubb}
\ee
where $f_{\alpha,j}$ is a fermionic annihilation operator destroying  a particle of flavor $\alpha$ at site $j$, satisfying  anticommutation relations $\{ f_{\alpha,i},f_{\alpha',j}^\dagger\} = \delta_{\alpha,\alpha'}\delta_{i,j}$,   $\sum_{\langle ij\rangle}$ indicates a sum over nearest neighbors, and $\alpha$ and $\alpha'$ are flavor indices that run from $1$ to $N$.  It is useful to note that the interaction term may be rewritten as $(U/2) \sum_{i,\alpha} \hat m_i(\hat m_i-1)$, defining $\hat m_i\equiv \sum_\alpha n_{\alpha,i}$ and $n_{\alpha,i}\equiv f_{\alpha,i}^\dagger f_{\alpha,i}$.

\textit{Atomic limit: experimental observables and interpretation in $T/U\ll1$ and $T/U \gg 1$ limits.---}First we give results in the atomic limit ($t=0$), the zero'th order term in the high temperature series expansion in $t/T$.  Throughout, we will present explicit analytic results for the free energy density $\mc F$, from which the other considered observables can be obtained by differentiating: the average filling and entropy per site are $\expec{m}=-\partial \mc F/\partial \mu$ and  $s= -\partial \mc F/\partial T$, respectively.

For the homogeneous system,
 the grand canonical free energy per site for $t=0$ is~\cite{cazalilla_ultracold_2009}
\be
\mc F_0 &=& -T \log z_0\label{eq:atomic-F}
\ee
with the on-site partition function
$
z_0 = \sum_{m=0}^N
C_m^N e^{-\beta\epsilon_0(m)},\label{eq:atomic-z}
$
where $\epsilon_0(m) \equiv (U/2) m(m-1) - \mu m$, $\beta \equiv 1/T$,  $k_B=\hbar=1$ throughout, and
 $C_{k}^N$
the binomial coefficient.
The average filling is
$m_0= \expec{m}_0$ where the expectation  $\expec{\mc O}_0$ of an operator $\mc O$ that depends only on the site filling is defined as the on-site expectation
$
\expec{\mc O}_0 \equiv\frac{1}{z_0}\sum_{m}\mc O
C_m^N
e^{-\beta \epsilon_0(m)}
 $
 and the entropy is $s_0 = \log z_0 +(1/T)\expec{\epsilon_0(m)}_0$.

It is illuminating to consider the observables in the $T\ll U$ and $T\gg U$ limits.  In the high temperature $T\gg \{U,\mu\}$ limit the on-site partition function is $z_0(T\gg U) = \sum_{m=0}^N
 C_m^N
= 2^N$, the density is $m(T\gg U) = N/2$, and the entropy density is $s(T\gg U)= N\log 2$. Interpreting this, pairs of fillings $m$ and $N-m$ are equally likely, for all $m$, so that the average  filling is $N/2$, and each flavor is equally likely to occupy or not occupy each site, giving $\log 2$ entropy per flavor per site. In contrast, in the low temperature MI limit defined by $t \ll T\ll U$ and $\mu \ne Um$ for all $m$, one term dominates $z_0$ so that the partition function is $z_0=
C_m^N
e^{-\beta \epsilon_0(m)}$, the density is  $m$  chosen to minimize $\epsilon_0(m)$, and the entropy density is
$s=\log (C_m^N) $.
Again, this  may be simply understood: the filling is fixed at $m$ and the entropy per site for the filling $m$ MI state in this regime is just the number of ways to choose $m$ of $N$ flavors to put on a site. Finally, when $t\ll T \ll U$, but $\mu= Um'$ --- the metal boundary separating the $m'$ and $m'+1$ MI's ---
the density is $m=(m'+1)/(1+1/N)$, and the entropy density is $s=\log \lp 
C_{m'}^N
+
C_{m'+1}^N\rp$, with interpretations analogous to the discussions above.

Cold atomic systems are confined in trapping potentials.  To include their effects,  we employ a Thomas-Fermi or local density approximation (LDA)~\cite{pethick_bose-einstein_2001}.  We take
the trapped system's properties at a point in space $\v{r}$ to be those of the homogenous system at a chemical potential $\mu(\v{r}) = \mu_0 - V(\v{r})$, with $V(\v{r})$ the trapping potential.  This is accurate when the potential varies slowly compared to the state's characteristic lengths, which is well satisfied in the regimes of present interest.  For simplicity, and as an accurate description of most traps, we approximate $V(r)=m\omega^2 r^2/2 $ where $m$ is the particle mass an $\omega$ is the external potential's trapping frequency.

Figure~\ref{fig:compare-t-t0-profiles} shows the density $m$ and entropy $s$ profiles,
 and the formation of Mott plateaus at low temperatures.  It also shows the effects of the tunneling corrections calculated later.
At high temperatures $T/U\gsim 0.2$ all curves are smooth and lack visible signs of Mott plateaux (not shown).  Fig.~\ref{fig:compare-t-t0-profiles}'s left and right insets show the $N$-dependence near $m=1$ and $m=N/2$, respectively.  Although the density profiles shown are at temperatures below the regime of validity $t/T\gsim 1$, one finds that the theory is inaccurate only near chemical potentials halfway between the Mott shells~\cite{jaerdens_quantitative_2010}. Working deeper in the approximation's regime of validity (smaller $T$) would lead to a smaller effect, invisible to the eye.  However, the qualitative effects are the same, only smaller.

\begin{figure}[hbtp]
\setlength{\unitlength}{1.0in}
\vspace{-0.1in}
\includegraphics[width=3.5in,angle=0]{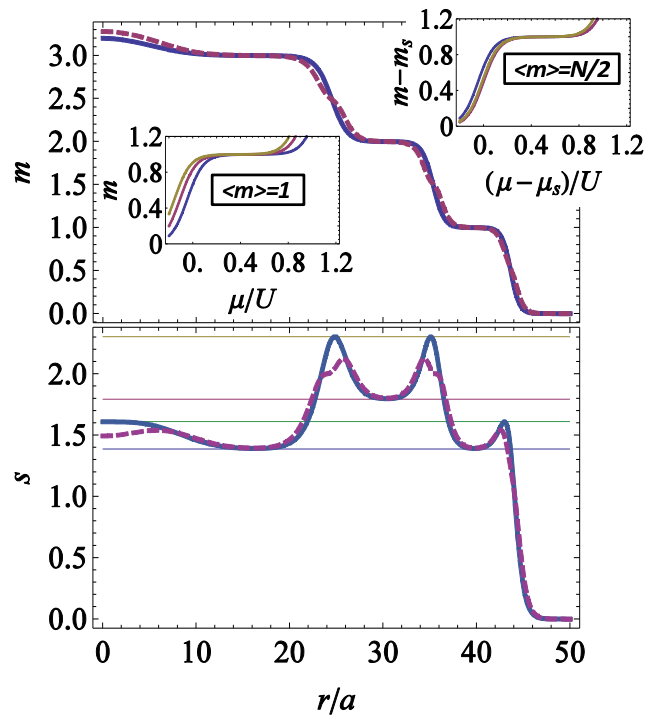}
\vspace{-0.45in}
\caption{Observables as a function of distance to trap center $r$  for $N=4$ at $T/U=1/15$, $\mu/U=3.0$, and in two dimensions. Top: density (solid: atomic limit; dashed: $t/U=0.1$).  Although the value of the $t/T$ is not deep in the regime where the high temperature expansion is valid, such a large value was chosen so the effects would be visible in the density profile.
Insets: density versus $\mu$ for $N=2,5,$ and $10$, showing $N$ and $m$-dependence.   Left inset: zoom around $m=1$ shell. Right inset: zoom around $m=N/2 $ with density and  $\mu$ shifted by $m_s$ and $\mu_s$ so that the Mott shells nearly overlap on the plot.
Bottom: entropy (solid: atomic limit; dashed: $t/U=0.04$; horizontal dashed: $T\ll U$ deep Mott and deep metal limits discussed in text).
\label{fig:compare-t-t0-profiles}}
\end{figure}

\textit{Adiabatic loading.---}We provide a theoretical description for the standard protocol used to realize both bosonic~\cite{rey_role_2006} and fermionic~\cite{esslinger_fermi-hubbard_2010,joerdens_mott_2008}
MI's.  This is essential to understand and optimize the process.
Remarkably, we find that for the relevant $N$, the final achievable temperature substantially \textit{decreases} with increasing $N$.

The procedure used to create MI's is to first create a degenerate, weakly interacting gas  without a lattice.  A lattice is then ramped up to its final value.  Ideally, the ramp is slow enough that the process is adiabatic.
The adiabatic limit appears to be approached in recent boson experiments~\cite{rey_role_2006,gemelke_situ_2009},
and with less than $\sim 50\%$ increases of entropy in some $SU(2)$ Fermi experiments~\cite{joerdens_mott_2008,esslinger_fermi-hubbard_2010}.
The  limits of adiabaticity are ill-understood and beyond the scope of our present work.

In the adiabatic limit entropy is conserved, and given the initial state's particle number and entropy one can determine the final temperature by matching the particle number and entropy to the initial state.  The initial state entropy may be controlled and measured only indirectly, however, through the temperature. Thus, 
we must first determine $S$ of the initial state from the initial temperature $T_1$.

For any initial state sufficiently cold to reach a MI, the initial gas will be deeply degenerate, $T\ll \mu$. For large numbers of particles, the harmonic trap can be treated as having a continuous density of states  $\nu(\epsilon)=(\epsilon/\omega)^{d-1}/(\omega(d-1)!) \Theta(\epsilon)$ with $\Theta(\epsilon)$ the Heaviside step function. Using this, the total particle number and entropy of a $d$-dimensional trapped system are
$
M = \frac{N}{d!}\lp\frac{\mu}{\omega}\rp^d$
and
$S = \frac{T}{\omega} \frac{N\pi^2}{3(d-1)!}\lp\frac{\mu}{\omega}\rp^{d-1}
$
to lowest order in $T/\mu$.  Note that to the same accuracy as this approximation, the Fermi temperature is $T_F=\mu$.

Figure~\ref{fig:TVsTi-constV0-slice}  overviews the results of our adiabatic loading calculation for a two dimensional system.  Although specific values depend on microscopic parameters such as the scattering length $a_s$, lattice spacing $a$, and trap length $\ell$, the qualitative findings discussed below are independent of these.  Fig.~\ref{fig:TVsTi-constV0-slice} shows the final temperature $T/U$ after adiabatic loading as a function of the initial temperature $T_i$ for several $N$, using experimentally relevant parameters at fixed final lattice depth. Fig.~\ref{fig:TVsTi-constV0-slice} (inset) illustrates the effect on density profiles of cooling obtained by increasing $N$.   For reference, the coldest alkaline earth gases in the weakly interacting regime are at roughly  $T/\mu=0.14$ for $^{173}$Yb ($N=6$) and $T/\mu=0.26$~\cite{taie_realization_2010} for $^{87}$Sr ($N=10$)~\cite{takasu_spin-singlet_2003}.  Also, we have found that in the deep lattice regime, the final rescaled temperature $T/U$ is independent of lattice depth $V_0$ within a few percent.


One of our most remarkable findings is the final temperature's $N$-dependence: increasing $N$ produces colder MI's with initial parameters fixed in an experimentally realistic way.  In particular, this result holds when
the initial temperature  of the weakly interacting gas is fixed, as would
hold if the gas were sympathetically cooled with another species to prepare the initial state --- an even more favorable situation is considered below.  A corollary is if adiabaticity can be maintained, since $SU(N)$ systems can be produced with initial temperatures comparable to current $SU(2)$ systems  MI's with $T\ll U$ and even $T\sim t$ are well within reach for $SU(N)$ fermion experiments.

To understand our findings, we consider the $N$-dependence of $S$ in the initial and final states. In the initial state, $S$ at fixed $M$ scales as $S_i\propto N^{1/d}$.  
For the MI at $T/U\ll 1$, the entropy scales as
$\log C_m^N$.  Considering the $m=1$ MI shell for simplicity,
$S_{\text{f}}\propto\log N$. Thus, naively one may expect the initial state's entropy to grow faster with $N$ than the final state's at fixed initial and final temperatures, but this is true only for very large $N$: in practice, for $d=3$, the power law grows slowly compared to the logarithm, and for $N\lsim 20$ the ratio of MI entropy to metal entropy \textit{increases} with $N$. This results in  cooler states with increasing $N$ in this regime, as observed in Figs.~\ref{fig:TVsTi-constV0-slice}. Similar effects persist in lower dimension, but to smaller $N$ (roughly 7 and 3 in $d=2$ and $1$, respectively).  For $m$ that scale with $N$, e.g. $m=N/2$, the situation is even more favorable: one finds that the MI entropy is proportional to $N$ and thus for any $d>1$ the final states get colder with increasing $N$ \textit{for all $N$}. A similar argument explains the reversal of this effect at higher temperatures (see Supplementary information).

While we observe this dramatic cooling when fixing $T$, as appropriate, say,  for sympathetic cooling, the actual situation may be even more favorable. Other cooling procedures may fix the initial gas's entropy or $T/\mu$ independent of $N$.  In this case the MI will get colder with increasing $N$ \textit{for all $N$}, even for  $m=1$. A simple candidate method is direct evaporative cooling of the fermions, with an added benefit that  Pauli blocking becomes less important as $N$ increases.

\begin{figure}[hbtp]
\setlength{\unitlength}{1.0in}
\includegraphics[width=3.4in,angle=0]{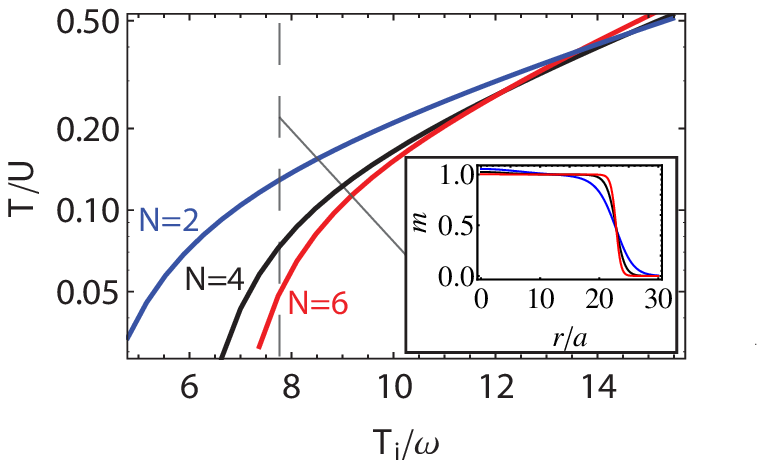}
\caption{
Left: Adiabatic loading: final temperature $T/U$ as a function of initial temperature  $T_i/\omega$ for $N=2,4,6$ (top to bottom at low temperatures) at lattice depth $V_0=10E_R$, with $E_R =\hbar^2 \pi^2/(2ma^2)$ the atom recoil energy, in two dimensions.  Parameters are fixed to experimental values for $^{173}$Yb\cite{taie_realization_2010}: we find   $U/\omega=50$, $a/\ell=0.4$, and $M=5\times10^4$ using standard optical lattice results for $U$~\cite{zwerger_mott-hubbard_2003}. The experimental initial gas temperature is  $T_i/\omega=5.2$ ($T_i/\mu=0.14$ for $N=6$).  Parameters are roughly the same for other alkaline earth atom experiments.
Right: Density profiles after adiabatic loading for $T_i/\omega=7.3$ ($T_i/\mu=0.14,0.17$, and $0.20$ for $N=2,4,6$, respectively).
\label{fig:TVsTi-constV0-slice}
}
\end{figure}



\textit{Non-zero tunneling.---}Here we incorporate
tunneling by a second order expansion in $t/T$, equivalent to a finite temperature $t/U$ expansion.
 Analogous results were recently applied to $SU(2)$ Fermi gases~\cite{jaerdens_quantitative_2010}, where this second order series is quantitatively accurate ($\lsim 1\%$ error)  when $T\gsim t$ for fillings in the metallic region (e.g. $m\sim 1.3$) and for much lower temperatures at other fillings (e.g., $m\sim 0, 1$, and $2$)~\cite{fuchs_thermodynamics_2010,de_leo_trapping_2008}. Present experiments are in this regime,
 although $SU(2)$ experiments have begun to reach temperatures $T\lsim t$.
 Longer series maintain agreement for $T\gsim t$, but, being divergent, series of all lengths give unphysical results for $T\lsim t$.  Longer series may nevertheless be useful in understanding low
temperature physics by analyzing their analytic structure~\cite{oitmaa_series_2010}.

We find the free energy density to $O[(t/T)^2]$ is
\be
\mc F &=& \mc F_0 -T z \lp\frac{t}{U}\rp^2\expec{m}_r\lb 1 - \frac{\expec{m l}_r}{N\expec{m}_r}\rb\label{eq:F-2nd-order}
\ee
where we define
\be
\begin{Bmatrix}\expec{m}_r \\ \expec{ml}_r
\end{Bmatrix}
&=& \frac{1}{z_0^2}\bigg [\frac{1}{2}\lp\frac{U}{T}\rp^2\sum_{m} \lp  C_m^N\rp^2 \begin{Bmatrix}
 m \\ m^2
 \end{Bmatrix} e^{-2\beta \epsilon_0(m)}\nonumber \\
 &&\hspace{0.18in}{}+ \sum_{m=0}^N\sum_{ \Large{\tworow{l=0,}{l\ne m}}}^N \begin{Bmatrix}
 m \\ ml
 \end{Bmatrix} g_{ml}\bigg ],
\ee
with
$g_{ml} \!\equiv
C_m^N C_l^N
\!\lb  \frac{e^{-\beta U(m-l+1)}+\beta U \lp m-l+1\rp-1}{(m-l+1)^2e^{\beta(\epsilon_0(m)+\epsilon_0(l))}}\rb\!\!.$
Eq.~\eqref{eq:F-2nd-order}'s bracketed term comes from Pauli blocking.

Figure~\ref{fig:compare-t-t0-profiles} shows tunneling's consequences for a two dimensional system. In the density profiles, we see that it reduces the MI's size, as expected.  In the entropy profiles,  tunneling slightly increases the MI entropy while significantly decreasing the metal entropy: in the MI, tunneling reduces the excitation gap, increasing the entropy, while in the metal, the increasing bandwidth lowers the low energy density of states, decreasing the entropy.  We also observe (Fig.~\ref{fig:compare-t-t0-profiles} insets) that although the \textit{magnitude} of the tunneling corrections depends only weakly on $m$, the $N$-dependence of the tunneling corrections depends more strongly. For $m=1$, there is significant $N$-dependence of the tunneling corrections, while there is little in the $m=N/2$ shell.

We finally have considered the effect of tunneling on the adiabatic loading procedure, and find that  tunneling increases the final temperature.  This may be understood by considering Fig.~\ref{fig:compare-t-t0-profiles}: tunneling reduces the total entropy in the trap as a consequence of a small increase in MI entropy and a relatively larger decrease in metal entropy.  The effects are small when $t/T$ is small, the  limit in which our calculations are valid.  For example, an $N=5$, $d=2$ system with $U/\omega=50$ and $a/\ell=0.13$ initially at $T/\mu=0.2$ adiabatically loads to a MI with temperature $T/U=0.12$ for $t/U=0$ and $T/U=0.14$ for $t/U=0.1$.  Tunneling corrections somewhat decrease the cooling effect of increasing $N$.

\textit{Conclusions and discussion.---}We studied the metal-Mott insulator crossover using a high-temperature series expansion technique, up to second order in $t/T$.  We calculated the density and entropy in the atomic limit and quantified how tunneling reduces the size of the Mott insulating region, its flavor number  ($N$) and filling ($m$) dependence, and how it increases the Mott entropy while decreasing the metal entropy.

Additionally, we studied the standard experimental protocol used to realize Mott insulators and showed that the final temperatures significantly decrease with increasing $N$. Our explicit calculations were in two dimensions, but similar results hold in other dimensions, becoming weaker in one dimension and more dramatic in three dimensions. This is an encouraging result for attempts to reach large-$N$ Mott insulators, which paves the way to study exotic physics.  The $N$-dependence of low temperature $T\ll t$ behavior --- e.g., antiferromagnets or other ordered or exotic states --- remains to be investigated.

Three body losses might limit the observability of the shell structure with $m>2$.  Nevertheless, based on Ref.~\cite{esry_recombination_1999}'s theory we find, at least for $^{87}$Sr, that even fillings $m\lsim 5$ may live sufficiently long to explore their many-body physics (see Supplementary material).

\textit{Acknowledgements.}---We thank Salvatore Manmana and Adrian Feiguin for discussions. AMR and KH  are supported by grants from the NSF (PFC and PIF-
0904017), the AFOSR, and a grant from the ARO with
funding from the DARPA-OLE. MH is supported by
DOE grant number de-sc0003910, and VG is supported
by NSF grant number DMR-0449521 and  PIF-0904017.


\clearpage

\title{Supplementary information for: High temperature thermodynamics of fermionic alkaline earth atoms in optical lattices}


\author{Kaden R.~A. Hazzard$^{1,2}$} \email{kh279@cornell.edu}
\author{Victor Gurarie$^2$}
\author{Michael Hermele$^2$}
\author{Ana Maria Rey$^{1,2,3}$}
\affiliation{$^{1}$ JILA, University of Colorado, Boulder, Colorado 80309-0440, USA}
\affiliation{$^{2}$ Department of Physics, University of Colorado, Boulder, Colorado 80309-0440, USA}
\affiliation{$^{3}$ NIST, Boulder, Colorado 80309-0440, USA}

\maketitle

\section{Other observables}

In the main text, we concentrated on the density and entropy.  This was largely because the density is the most common observable, while the entropy controls the temperature attained during adiabatic loading.  However, other interesting quantities may be measured.

\begin{figure}[htbp]
\setlength{\unitlength}{1.0in}
\includegraphics[width=3.3in,angle=0]{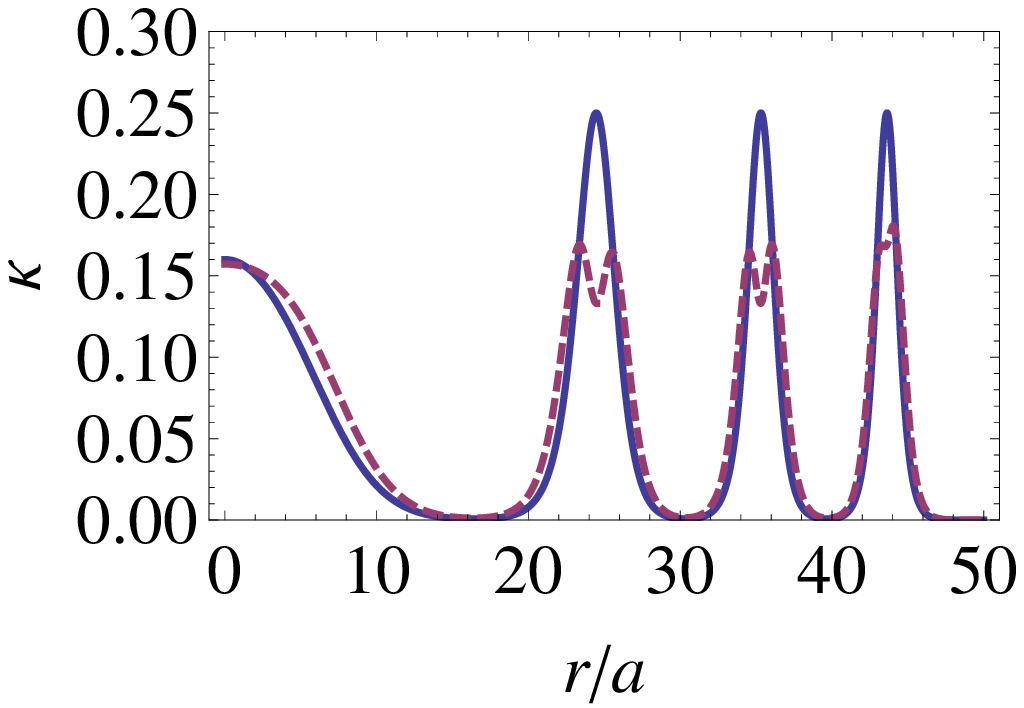}
\includegraphics[width=3.3in,angle=0]{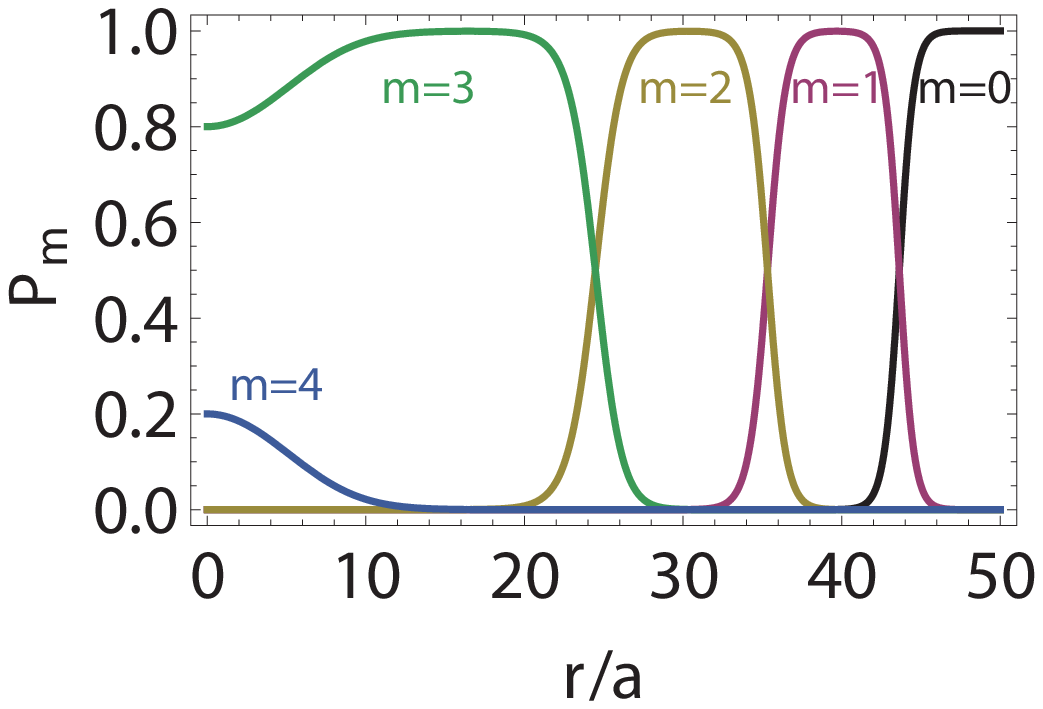}
\caption{Top: compressibility $\kappa$ as a function of distance to trap center $r$ for $N=4$ at $T/U=1/15$, $\mu/U=3.0$ (solid: atomic limit; dashed: $t/U=0.1$).  These parameters correspond to the density profile shown in the main text.  The dips in the compressibility at the metal center are artifacts of applying our approximation for a $t/U$ that is larger than where the approximation remains valid.  This was to ensure that the effect of tunneling would be visibile in the density profile, in the main text.
Bottom: atom number distribution showing $P_1$, $P_2$, $P_3$, and $P_4$ for the same parameters as the compressibility, in the atomic limit.
\label{fig:other-observables}}
\end{figure}

Here, we discuss the compressibility $\kappa=\partial n/\partial \mu= -\partial^2 \mc F/\partial \mu^2$ and the on-site atom number distribution $P_m$, the probability of $m$ particles occuring on a site.  The former is simply a derivative of the density (or the free energy) and is thus readily available from our calculation. The latter is \textit{not} simply a derivative of the free energy we compute, without adding extra terms in the $SU(N)$ Hamiltonian, and so we only consider it in the atomic limit.  In this limit, one obtains $P_m =(1/z_0) C_m^N e^{-\beta \epsilon_0(m)}$ using the same notation as in the main text: $\epsilon_0(m)=(U/2)m(m-1)-\mu m$ and $z_0=\sum_{m=0}^N  C_m^N e^{-\beta \epsilon_0(m)}$.
It is worth noting that once one knows $P_m$, one knows all moments of the on-site density: $\expec{m}$, $\expec{m^2}$, $\ldots$  However, note that although for the $SU(2)$ case the double occupancy is uniquely determined once one knows the density and number fluctuations $\expec{n^2}-\expec{n}^2$, the same is not true for the $SU(N)$ gas.

Figure~\ref{fig:other-observables} shows $\kappa$ and $P_m$ as a function of the distance to trap center for the same parameters as Fig.~1 (bottom) of the main text.  For these parameters, the compressibility is nearly zero in the Mott regions and peaks in the metal as expected.  Tunneling reduces the magnitudes of the peaks and slightly increases the magnitude of the compressibility in the Mott insulator as expected.  At larger temperatures $T\gsim 0.3 U$, this structure in the compressibility becomes quite washed out.  The atomic limit number distribution shows that in Mott regions, a single filling dominates the number distribution, while in the metal region there are two relevant fillings whose fillings change with $r$ to smoothly connect the Mott regions.  At higher temperatures $T\gsim 0.3U$, the Mott regions will have significant probability to occupy fillings other than the Mott filling and the metallic region will have significant probability to occupy fillings other than the two fillings that are relevant at low temperature.  At sufficiently high temperature, all of the structure is washed out.

\section{Why, at high temperatures, heats rather than cools with increasing $N$}

Figure~2 of the main text shows that at low temperature, the effect of increasing $N$ for experimentally relevant $N$ while fixing the initial temperature is to generate colder clouds.  However, it also shows that the trend is reversed at higher temperatures.  This may be somewhat counterintuitive, as at very high temperatures the entropy should become very high even in the deep lattice.
\begin{figure}[hbtp]
\setlength{\unitlength}{1.0in}
\includegraphics[width=3.4in,angle=0]{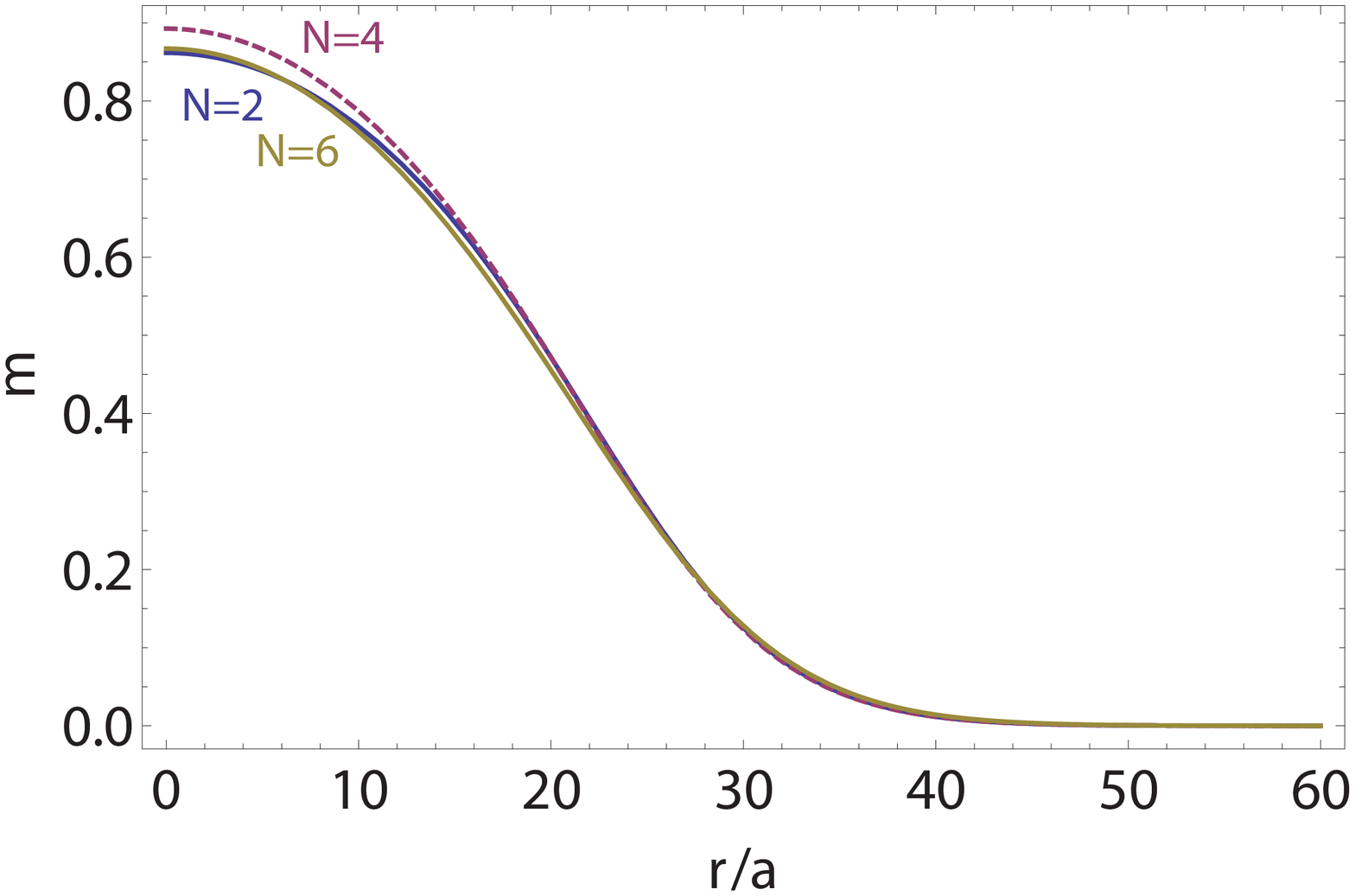}
\includegraphics[width=3.4in,angle=0]{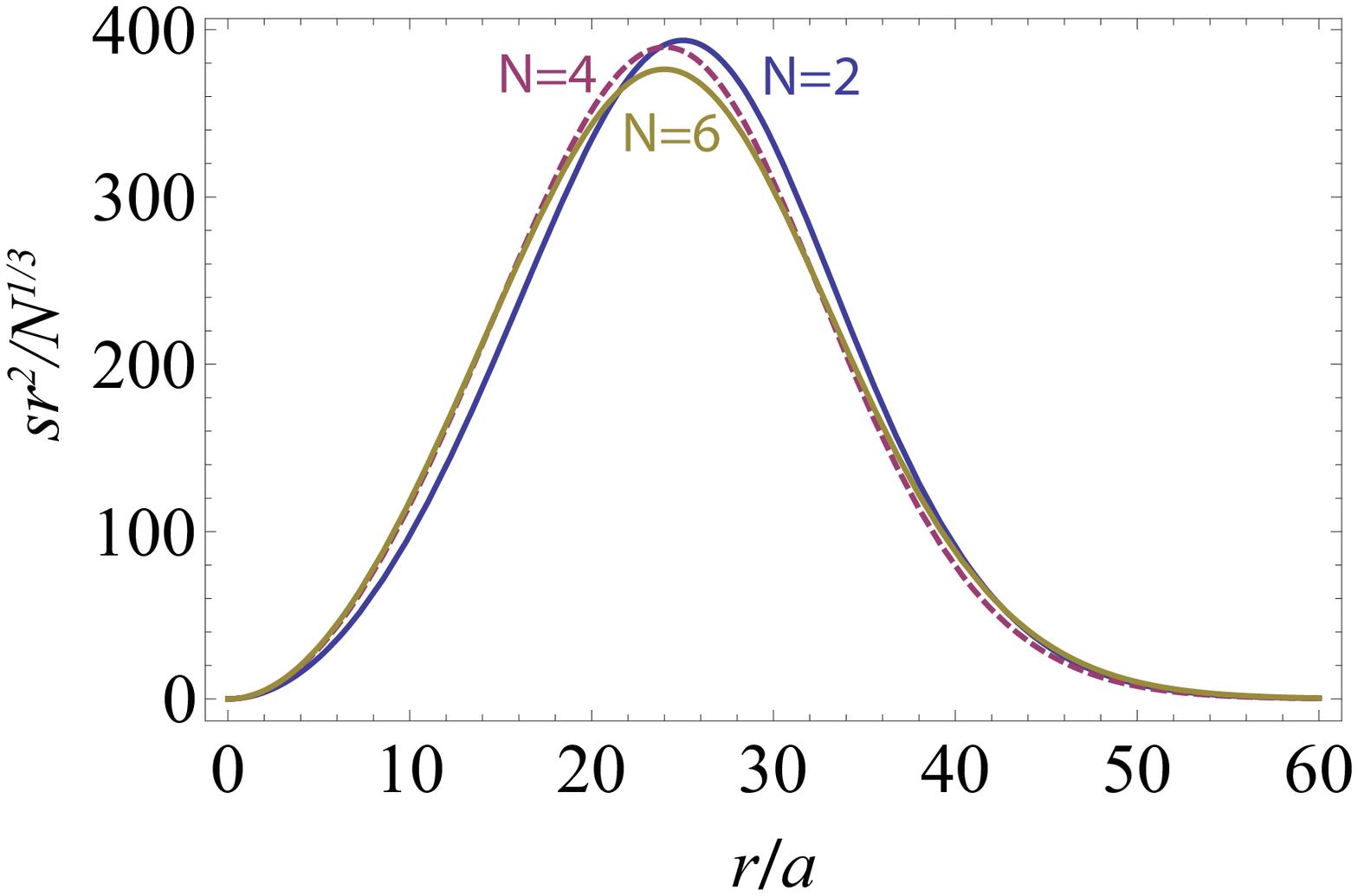}
\caption{Top: density profiles for $N=2,4,6$ (bottom curves are $N=2$ and $N=6$, colored blue and gold, respectively, and are extremely similar; top curve is $N=4$.) of final states after adiabatic loading for $T_i/\omega=14.7$ and other parameters corresponding to Fig.~2 of the main text.
Note the absence of a Mott regime at these high final temperatures.
Bottom: entropy profiles for $N=2,4,6$ (top/blue, middle/dashed/purple, bottom/gold respectively), rescaled by $r^2 /N^{1/3}$, of final states after adiabatic loading for $T_i/\omega=14.7$ showing that the entropy scales as $N^{1/3}$ at these high temperatures rather than the more favorable $\log N$ obtained at lower temperatures.  (The $r^2$ is the surface area factor that multiplies the entropy density in the integral to obtain the total entropy, and is included simply so one may more easily see what regions of the system contribute most to the total entropy.)
\label{fig:high-temp-heating}}
\end{figure}

To understand this behavior, we first note that at the higher temperatures the density profiles are quite different than at low temperatures: they have lower central density and are more spread out.  Fig.\ref{fig:high-temp-heating} illustrates this effect.  In particular, there is no central Mott regime in the final states and it turns out this reduces the entropy's dependence on $N$.  In particular, we recall that at low temperatures, the final state entropy grows with $\log N$, while the initial state grew with $N^{1/3}$ in three dimensions; for the $N\lsim 20$ of interest, the former grew faster than the latter, leading to the cooling.  In contrast, Fig.~\ref{fig:high-temp-heating} illustrates that with the lower density clouds produced at high temperature, there is no Mott region and the entropy always scales roughly as $N^{1/3}$ rather than $\log N$, eliminating the cooling effect seen at low temperatures.  The details of this heating will depend on the central filling, and will presumably usually occur at higher temperatures for higher central fillings.

\section{Loss rates}

As a final note we mention particle losses.
The dominant loss source will be three-body losses with recombination rate $\gamma \expec{m(m-1)(m-2)}$.  For filling $m=1$ and $m=2$  Mott insulators, the lifetimes will be quite long, limited only by losses from number fluctuations to higher fillings for all alkaline earths.  For larger fillings,
we estimate loss rates based on the theory of Ref.~\cite{esry_recombination_1999}, averaging interference of the two main recombination pathways.  This is accurate to a few percent for both small $a$ and
 large $a$
  and overestimates losses in between.  For the $^{87}$Sr experiments, we find $\gamma =0.03 s^{-1}$ in a very deep $V_0= 35E_R$ lattice, with slower loss rates in weaker lattices.  Correspondingly, the $m=3 $ shell in the atomic limit (no number fluctuations) has a lifetime $\tau_3\sim6$s.   For $^{173}$Yb experiments, due to the roughly doubled scattering length we find six times faster loss rates at the same lattice depth (in recoil units) $V_0/E_R$  and the same lattice spacing as in the previous ${}^{87}$Sr estimates.


However, this comparison is complicated by three points.   Firstly, the energy scales in the Hubbard Hamiltonian  also increase, by a factor of about two due to the doubled scattering length.  Hence, the relevant quantity --- the ratio of the loss rates to those energy scales --- is only larger for $^{173}$Yb by a factor of three fixing $V_0/E_R$, working at the same lattice depth and spacing.

Unfortunately, there is a second consideration which makes things much less favorable for \textit{current} experiments. Current $^{173}$Yb experiments are performed in lattices with wavelength shorter by a factor of about two, which increases the loss rates enormously ($\sim 2^6$). This factor comes because, for fixed $V_0/E_R$, halving the lattice spacing in three dimensions increases the density by $2^3$ and the on-site losses go like maximum density squared (an integral over space of the density cubed $\int d\v{r} \rho^3(\v{r})$). This is somewhat compensated by an increasing Hamiltonian energy scales.  The interaction increases proportional to the maximum density ($\sim \int d\v{r} \rho^2(\v{r})$), by a factor of $\sim 2^3$.  This gives an overall loss rate, relative to rates governed by the Hamiltonian, a factor of $8$ larger than the estimate above, at the same $V_0/E_R$.  This still may be tolerable; the $m=3$ Mott insulator has a lifetime of about $40$ms and in deep lattices the lower shells have very long lifetimes.
Furthermore, this constraint to use a much shorter lattice spacing is not fundamental.

Moreover, a third consideration shows that these estimates are somewhat too pessimistic.  We assumed that $V_0/E_R$ was fixed, but both the increase in scattering length $a_s$ and decrease in lattice spacing $a$ decrease the  $V_0/E_R$ required to obtain the same $t/U$, thus counterbalancing the increased loss rates somewhat.
The details of the balance of these effects depend on the regime of the phase diagram being explored.



\end{document}